# INWARD BOUND: STUDYING THE GALACTIC CENTER WITH NAOS/CONICA

T.Ott[1], R.Schödel[1], R.Genzel[1,2], A.Eckart[3], F.Lacombe[4], D.Rouan[4], R.Hofmann[1], M.Lehnert[1], T.Alexander[5], A.Sternberg[6], M.Reid[7], W. Brandner[8,9], R. Lenzen[8], M. Hartung[8], E.Gendron[4], Y.Clénet[4], P.Léna[4], G. Rousset[10], A-M. Lagrange[11], N. Ageorges[9], N. Hubin[9], C. Lidman[9], A.F.M. Moorwood[9], A.Renzini[9], J. Spyromilio[9], L.E.Tacconi-Garman[9], K.M. Menten[12], N.Mouawad[3]

[1]Max-Planck-Institut für extraterrestrische Physik (MPE), Garching, Germany
[2]Dept. of Physics, University of California, Berkeley, USA
[3]1.Physikalisches Institut, Universität Köln, Germany
[4] Observatoire de Paris-Meudon, France
[5] Faculty of Physics, Weizmann Institute of Science, Rehovot, Israel
[6] Department of Astronomy & Astrophysics, Tel Aviv University, Israel
[7] Center for Astrophysics, Cambridge, USA
[8] Max-Planck-Institut für Astronomie, Heidelberg, Germany
[9] European Southern Observatory, Garching, Germany
[10]Office National d'Etudes et de Recherches Aèrospatiales (ONERA), Chatillon, France
[11]Observatoire de Grenoble, Grenoble, France
[12] Max-Planck-Institut für Radioastronomie, Bonn, Germany

## INTRODUCTION

Because of its proximity[1] the Center of the Milky Way is a unique laboratory for studying physical processes that are thought to occur generally in galactic nuclei. The central parsec of our Galaxy is known to contain a dense and luminous star cluster, as well as several components of neutral, ionized and extremely hot gas. For two decades, evidence has been mounting that the Galactic Center harbors a concentration of dark mass associated with the compact radio source SgrA* (diameter about 10 light minutes), located at the center of the nuclear star cluster. SgrA* thus may be a supermassive black hole. High resolution observations offer the unique opportunities of stringently testing the black hole paradigm and of studying stars and gas in the immediate vicinity of a black hole, at a level of detail that will never be accessible in any other galactic nucleus.

At visible wavelengths, the Galactic center is hidden behind 30 magnitudes of Galactic foreground extinction. Optical and ultraviolet observations are impossible or exceedingly difficult. Our knowledge of the Milky Way center thus derives from observations at both shorter (X- and γ-rays) and longer wavelengths (infrared and radio) that can penetrate through this obscuration. Near-infrared observations from large ground-based telescopes are particularly useful for studying the stellar component. Stars are excellent test particles of gravity. Stellar motions are ideal tools for significantly improving the constraints on the concentration and location of the central mass inferred from earlier observations of gas velocities. For this purpose, over the past decade a group of us at the Max-Planck-Institut für extraterrestrische Physik (MPE) has been using the high resolution, near-infrared camera SHARP, employed as a regular visiting instrument on the 3.5m ESO NTT. The primary goal of the NTT observations was to detect stellar proper motions in the immediate vicinity of SgrA* by exploiting the excellent imaging properties and seeing of that telescope for diffraction limited H- and K-band, speckle imaging of the Galactic Center. A second experiment used the novel '3D' near-IR integral field spectrometer for measuring Doppler velocities on the 2.2m ESO/MPIA telescope.

---

[1] We assume here that the Galactic Center is 8 kpc from the Sun. At that distance 1 arcsecond corresponds to 0.038 pc, or $1.2 \times 10^{17}$ cm

When the first SHARP and 3D results of the stellar dynamics emerged 6 years ago (Eckart & Genzel 1996, 1997, Genzel et al. 1996, 1997) they provided a compelling case for the presence of a compact, central mass. The stellar velocity dispersion increases towards SgrA* from 20" to <1" with a Keplerian law ($\sigma_v \propto R^{-1/2}$). Several stars in the 'SgrA* cluster' within 1" of the radio source exhibit velocities in excess of $10^3$ km/s. Shortly thereafter, the NTT results were confirmed and strengthened by independent proper motion data obtained with the 10m Keck telescope (Ghez et al. 1998). The Keck group also reported two years later the first orbital accelerations of three stars, demonstrating that the center of force was indeed centered on the radio source (Ghez et al. 2000). The NTT and Keck results proved that 2-3.3x$10^6$ $M_\odot$ are concentrated within ≤10 milli-parsec (2000 AU) of SgrA*. This mass concentration is most likely a supermassive black hole.

## NAOS/CONICA Observations

While a central black hole is the most probable configuration, the NTT/Keck data before 2002 could not exclude several alternatives, such as an extremely compact cluster of dark astrophysical objects (e.g. neutron stars, stellar black holes, or white dwarfs), or a degenerate ball of hypothetical heavy neutrinos. Even if most of the mass is in a black hole, the hole may be surrounded by a halo of heavy stellar remnants or heavy particles, or may even not be a single object. To address these questions, it is necessary to determine the gravitational potential closer to SgrA*, and with greater precision than was previously possible. The resolution of an 8m-class telescope is required. It is desirable to replace the statistical method of deducing the mass distribution from averaging velocities of many stars by the more precise technique of determining masses from individual stellar orbits. There are also a number of other intriguing issues that could not be addressed by the comparatively shallow NTT/Keck observations. If SgrA* is indeed a massive black hole there should be a cusp of low mass stars centered on it. It would also be important to explore other theoretical predictions characteristic of the black hole environment, such as stellar collisions and disruptions, or gravitational lensing. These goals call for much more sensitive measurements than have been achievable with speckle techniques. Adaptive optics (AO) imaging with good Strehl ratios is required. It is of great interest to better understand the evolution of the nuclear star cluster itself. Given the existing evidence for hot, massive stars in the central parsec (e.g. Genzel et al. 1996) one would like to understand how they have come into being in the very dense central environment. Finally it is highly desirable to detect the infrared counterpart of SgrA* itself. Infrared flux, spectral energy distribution and polarization provide important constraints on the poorly understood properties of black hole accretion flows.

The ideal instrument for tackling all these basic questions is the new AO imager/spectrometer NAOS/CONICA. NAOS/CONICA was mounted on UT4 (Yepun) in late 2001, and first observations became possible in early/mid 2002 during the commissioning and science verification phases of the instrument. Compared to the AO imagers at the Keck and Gemini telescopes, NAOS/CONICA has two unique advantages. First, the Galactic center passes through zenith, compared to a minimum airmass of 1.5 at Mauna Kea. Second, NAOS/CONICA has an infrared wavefront sensor. This permits wave front correction on the bright red supergiant IRS7 located 5.5" north of SgrA*. All other instruments have to lock on a relatively faint visible star at a distance of about 20". Furthermore, the $1K^2$ CONICA detector gives a field of view as large as 30", resulting in a major improvement in the astrometric analysis of the infrared data (see below). After discussion and consultation with the ESO Director General, it was decided, therefore, to immediately carry out a first series of imaging observations of the Galactic Center with the new instrument, as part of a collaboration of the two instrument teams and the ESO/Paranal staff. Beyond the scientific results, it also turned out that the Galactic Center was an optimal target for verifying and optimizing the technical performance of NAOS/CONICA. The 2002 NAOS/CONICA observations immediately gave us the best Galactic Center images ever taken, and led to a major breakthrough in our knowledge of the central parsec. The data are publicly available in the ESO archive. We give here a first account of what we have learned (see also Schödel et al. 2002, 2003, Ott et al. 2003, Genzel et al. 2003, Reid et al. 2003).

## Image Analysis and Astrometry

The observations were carried out during commissioning and science verification with NAOS/CONICA between March and August 2002. Data were taken in H-band, (1.65µm), $K_s$-band (2.16µm) and L'-band (3.76µm) with the 13 and 26 milli-arcsec (mas) pixel scales, with seeing varying between 0.5 and 1.3". For H and K we used the infrared wavefront sensor of NAOS to close the loop on the bright supergiant IRS 7 ~5.5″ north of SgrA*. At L' we closed the AO loop on a V~14 star ~20″ to the north-east of SgrA* (at the time of the observations the dichroic allowing the simultaneous use of the infrared wavefront sensor and the L/M-bands was not available). The individual images were flat-fielded, sky-subtracted and corrected for dead/bad pixels. The final frames were co-added with a simple shift-and-add (SSA) algorithm. In Figure 1 we show the $K_s$-band SSA image (left inset), as well as an H/$K_s$/L'-color composite, SSA image (right inset) obtained as part of the science verification run in August 2002. These are the best images of the 2002 season, with Strehl ratios >50% at $K_s$ and 33% at H. With integration times of about 20 minutes in each band, the data reach to H/$K_s$~20 and L'~14, ~3 magnitudes deeper than any Galactic center image taken before. The dynamic range of the images is about 13 magnitudes but bright stars are saturated in the deepest images. The final diffraction limited resolution was 40, 55 and 95 milliarcseconds (mas) in the H, $K_s$ and L'-bands (FWHM). The images in Figure 1 demonstrate the complexity of the dense stellar environment in the central parsec. Bright blue supergiants (in the IRS16 and IRS13 complexes), as well as red supergiants (IRS 7) and asymptotic giant branch stars (IRS12N, 10EE and 15NE) dominate the H- and $K_s$-images. At L' there is an additional group of dusty sources (IRS1, 3, 21). Extended L' emission comes from hot dust in the gaseous 'mini-spiral' streamers comprising the most prominent features of the SgrA West HII region. The immediate vicinity of SgrA* lacks bright stars and dust. There is a concentration of moderately bright ($K_s$~14) blue stars centered on the radio source (the 'SgrA* cluster'). The faintest sources recognizable on the images are equivalent to ~2 $M_\odot$ A5/F0 main sequence stars.

It is a highly challenging undertaking to extract photometry, source counts, stellar density distribution and astrometry in such an extremely dense, high dynamic range stellar field, and faced with the complex and time variable, point spread function (PSF) delivered by adaptive optics. Our experience from a decade of SHARP data analysis is that the best method for disentangling sources, reducing the influence of the seeing halos of the numerous bright stars and determining fluxes and accurate relative positions is to first deconvolve the images with one or several, well known methods (CLEAN, linear/Wiener filtering, Lucy/Richardson), prior to number counts and photometric analysis. The purpose of the deconvolution is not to enhance resolution, but to remove ('CLEAN') from the images from the effects of the AO PSF. For this purpose a mean PSF is extracted from several isolated and moderately bright stars across the field. In the case of Lucy and 'CLEAN' the final 'δ-function' images are re-convolved with a Gaussian of the original diffraction limited FWHM resolution. The agreement of sources identified with the different image analysis techniques is generally very good but naturally deteriorates within ~0.5" of the brightest stars (and in particular those which are saturated on our images). In these regions, graininess of the seeing halo, ringing and effects of the saturated central pixels make source identification of stars four or more magnitudes fainter than the bright star unreliable. We then identified point sources and carried out photometry with the FIND procedure from the IDL Astrolib library. We took the conservative approach of including only those sources that were present on both the Lucy/Richardson and Wiener deconvolved images, and also both in H and $K_s$. The final source lists comprise between 3200 and 4000 stars, depending on the FIND extraction parameters. To improve the reliability of the photometry, we applied different photometric algorithms and, where possible, averaged results from independent data sets. The final relative photometric errors are ≤0.1 mag below H=19, $K_s$=18 and L'=13, but probably twice that for fainter stars. The absolute photometry is uncertain to 0.15 mag in H and $K_s$, and 0.3 mag at L'. The PSF shape can vary significantly across a field as large as that shown in Figure 1, which spans a significant fraction of the isoplanatic angle (~20" in radius). Remarkably the anisoplanaticity is relatively small for the data shown in Fig.1, indicating that the seeing at the time of the observations was probably dominated by the surface layers. We have also found no significant impact of the anisoplanaticity on the relative positions, at least within the central 10" and to a precision of ~5mas (1/10 of the diffraction limited beam, Reid et al. 2003).

We determined incompleteness corrections for our images with the well known technique of first inserting and then again recovering artificial stars randomly across the entire field. Completeness maps were created by dividing smoothed maps containing the recovered artificial sources by smoothed maps with the initially added artificial stars. The maps thus designate the probability of recovering a source with a given magnitude at a given position. The average completeness of our $K_s$-images is 79%, 63% and 33% at $K_s$=17, 18 and 19, respectively. However, the completeness varies strongly with position, and can be very low near the brightest stars (e.g. in the IRS16 complex). All final source counts and faint star density distributions were corrected for these incompleteness effects.

An important improvement of the new NAOS/CONICA images is the accurate registration of the infrared images, in which the stars are observed, on the astrometric radio images, in which SgrA* is observed. For this purpose we aligned our NAOS/CONICA images with an astrometric grid using all 7 SiO maser sources in the field of view whose positions are known through interferometric radio measurements with the VLA and the VLBA with accuracies of a few mas (Reid et al. 2003, circled in Fig. 1). The SiO masers originate in the central 10 AU (~1 mas) of the circumstellar envelopes of bright red giants and supergiants, which are also present on the infrared images. We were able to improve the radio-to-infrared relative registration by a factor of 3 to ±10 mas from the larger number of stars with SiO masers in the CONICA images compared our earlier SHARP/NTT images. In our earlier work we had only used two SiO sources for the astrometric registration. That only allowed solving for center position, rotation angle and a single pixel scale of the infrared images. Our new analysis allows solving, in addition, for second order imaging terms. The nonlinear terms are negligible for NAOS/CONICA, but are significant for the SHARP data. Once the concurrent infrared and radio frames were aligned at their common observed epoch, we used the mean positions of more than a hundred stars to bring all other observed epochs into alignment by minimizing the residual astrometric errors. We were thus able to compute stellar positions (as offsets from SgrA* in right ascension and declination) with an accuracy of ±10 mas for all epochs between 1992 and 2002. From our 10 year, SHARP-NAOS/CONICA data set we are able to derive proper motions for about 1000 stars in the central 10" (Ott et al. 2003). In addition near-IR spectra are available for several hundred stars (Genzel et al. 1996, 2000, Ott et al. 2003). Finally we were able to determine individual stellar orbits (or orbital parameters) for 6 stars in the SgrA*-cluster (Schödel et al. 2002, 2003, Ghez et al. 2000, 2003, Eckart et al. 2002).

## Properties of the nuclear star cluster: Cusp and Young, Massive Stars

The images and spectra show that the nuclear star cluster in the central parsec is highly dynamic and rapidly evolving, contrary to expectations. In particular, there appear to be many massive stars that must have formed recently. We summarize here these new results and refer to Genzel et al. (2003) for a more detailed account.

Figure 2 (left) is a plot of the stellar surface density distribution for faint stars as a function of p, the projected separation from SgrA*. The K≤15 star counts for p(SgrA*)≥ a few arcseconds can be reasonably well fit by an isothermal model of core radius 0.34 pc. However, within a few arcseconds of SgrA* the NAOS/CONICA data clearly indicate an excess of faint stars above that of the flat isothermal core whose density increases toward the center. The two dimensional distribution (Fig2. (right inset)) shows that this cusp of faint stars is centered on SgrA*, within an uncertainty of ±0.2". This is in contrast to the near-IR light distribution (Fig.1), which is centered on the bright stars in the IRS16 complex. From a broken power-law analysis we find that the stellar density increases proportional to $R^{-1.37\pm0.15}$ within ~10" of SgrA*. The stellar density reaches about $10^8$ $M_\odot$ $pc^{-3}$ within ~0.5" of the radio source. SgrA* thus is at the center of the distribution of the faint stars. At first glance this result appears to be in very good agreement with theoretical expectations. The models predict the existence of such a cusp with power-law slopes ranging from 0.5 to 2, depending on the cusp's formation scenario and on the importance of inelastic stellar collisions. However, these models also predict that the stellar cusp ought to be dynamically relaxed and consist mostly of old, low mass stars. This is because stars migrate toward the central black hole by two-body relaxation processes. In the Galactic Center the relaxation time scale is about 150 Myrs for a ten solar mass star, and scales inversely proportional to mass. The relaxation time scale thus is much longer than the lifetime of such a star, unless its mass is less than about 2.5 $M_\odot$. We will see in the following that this expectation is not borne out by the observations. The data show a population of bright, apparently young stars that is not dynamically relaxed.

The central parsec contains two dozen or so, very luminous ($10^{5..6}$ $L_\odot$), blue supergiants. These 'HeI emission line stars' probably are massive (30-100 $M_\odot$), hot (20,000-30,000 K) stars in a short-lived, post-main sequence 'wind' phase, akin to late type (W{N,C}8/9) Wolf-Rayet stars and luminous blue variables (LBVs). They can account for most or all of the far-infrared and Lyman-continuum luminosity of the central parsec. Owing to their short lifetime, these stars must have formed in the last few Myr. Consistent with a recent formation time their proper motions are not relaxed, and are dominated by a turbulent rotation pattern, with an angular momentum opposite to that of Galactic rotation (Genzel et al.1996, 2000). About half a dozen of these luminous stars are located in the IRS16 and IRS13 complexes.

The new proper motions now demonstrate that also the somewhat fainter early type stars (K≤15) are not relaxed, and hence must be young as well. This is shown in Figure 3. While the old, late-type stars show a well mixed distribution of projected angular momenta, the early type stars exhibit a predominance of tangential orbits, that is, orbits with a velocity vector on the sky roughly perpendicular to the projected radius vector from the center. About 75% of all early type stars are on projected tangential orbits. In the central cusp (p≤3") about 60% of the K≤15 early type stars are on clockwise, tangential orbits. At the same time, the fraction of K≤15, late type stars decreases from

50% at p≥5" to <25% at p≤3". Un-relaxed early type stars thus dominate the counts of the cusp at moderately bright magnitudes. Further the K-band luminosity function (KLF) constructed from the new CONICA/NAOS data also changes toward the center. The Galactic Center KLF between $K_s$=8 and 19 is well described by a power law of slope ~0.21, but the p≤9" KLF in addition exhibits a prominent excess of counts near $K_s$=16. This bump is characteristic of old, metal rich and low mass (0.6-3 $M_\odot$), horizontal branch/red clump stars. This bump disappears in the central few arcseconds, indicating that the cusp lacks such old, low mass stars. Finally, speckle spectro-photometry and adaptive optics spectroscopy indicate that many of the fast moving 'SgrA* cluster' stars are early type as well, equivalent to late O, early B stars with masses of 15-20 $M_\odot$ (Genzel et al. 1997, Gezari et al. 2002, Ghez et al. 2003).

The observations portray a fascinating, albeit complicated picture of the nuclear star cluster. The central cusp appears not to be dominated by old, low mass stars. The presence of the many un-relaxed, massive stars suggests a highly dynamic picture. For the HeI emission line stars and the other early stars in the IRS16/13 complexes there are two plausible explanations. One postulates episodic infall and compression of very dense gas clouds, followed by in situ formation of stars. The other is the rapid sinking toward the center by dynamic friction of a massive young star cluster that originally formed at 5-10 pc from the center. Both mechanisms appear feasible but require very special, and somewhat unlikely conditions.

For the innermost early type stars in the SgrA*-cluster both of these scenarios fail. Instead, and owing to the very high stellar densities estimated above, continuous formation of moderately massive stars by collisions and mergers of lower mass stars appears to be the most plausible mechanism. The SgrA* cluster stars thus may be blue stragglers. The stellar collisional model is also supported by the simultaneous disappearance of the late type giants and the lower mass horizontal branch/red clump stars in the central cusp.

## A Star in a 15 Year Orbit Around SgrA*

Without any doubt the most exciting aspect of the new NAOS/CONICA data has been the first determination of stellar orbits around SgrA* (Schödel et al. 2002, 2003). The initial measurements of the orbital accelerations of S1 and S2, the two stars closest to SgrA*, were consistent with bound Keplerian orbits around a 3 million solar mass central object but still allowed a wide range of orbital parameters (Ghez et al. 2000). Specifically, possible orbital periods for S2 ranged from 15 to 500 years. When we obtained the first NAOS/CONICA images in March 2002, it immediately became obvious that S2 had moved to within10 mas of the radio source, and was now located east of SgrA*, that is, on the opposite side when compared to its position in the preceding years (Figure 4, left inset). This implied that S2 was just passing through its peri-center approach and that the S2 orbital data might probe the gravitational potential at a radius of about one light day, 20 to 30 times further in than our earlier, statistical determinations. When more data came in, this exciting interpretation was borne out. By late May 2002 it was clear that the SHARP and NAOS/CONICA positional measurements of S2 between 1992 and 2002 sampled two thirds of a highly elliptical orbit around the astrometrically determined position of SgrA*(Reid et al. 2003).

S2 orbits the compact radio source as the planets orbit our Sun. The data shown in the right inset of Figure 4 constrain all orbital parameters with better than 10% accuracy (Schödel et al. 2002, 2003). At its peri-approach in April 2002, the star was 17 light hours from SgrA* at which point it moved with ~8000 km/s. The rapid movement of the star could also be directly verified from the positional changes from month to month as more images were acquired (Fig.4). From the inferred orbital period of 15.2 years and its semi-major axis of 4.4mpc, Kepler's third law implies an enclosed mass of $3.3 \times 10^6$ $M_\odot$. This value is in excellent agreement with all earlier mass measurements between ~15 light days and several light years from the center (Figure 5). Despite its close approach to the central mass, S2 seems to have survived this encounter without any problems. The peri-center distance radius of S2 is 70 times greater than the distance from the black hole where the star would be disrupted by tidal forces (about 16 light minutes for a $17 M_\odot$). Since tidal energy deposition falls faster than the sixth power of the ratio of tidal radius to orbital radius, tidal effects are expected to be negligible, consistent with its lack of infrared variability.

Schödel et al. (2003) have recently reanalyzed the entire 1992-2002 SHARP-NAOS/CONICA image data set. In this analysis, initial positions and fluxes of a model cluster were first fitted to the best (August 2002) NAOS/CONICA image. Subsequently, positions at other epochs were obtained by using the preceding epoch as an initial guess. By iteratively fitting and subtracting from the images first the brighter and then ever fainter stars this analysis allows tracing the positions of the fainter stars even at epochs when they move close to bright stars. Ten isolated reference stars served as anchors to align center position and rotation angle for each epoch. By a rigorous selection of the very best imaging data and intensive image processing Schödel et al. obtained, high resolution maps of the central 1.5" around SgrA* down to K~16. They find six stars in this region that show clear signs of accelerations and allow determining/constraining their orbital parameters (Figure 6). In addition to S2, good orbital fits are also obtained for S3 and, to a lesser extent, for S0-16 (see below), S1 and S8. The resulting mass constraints are shown in Figure 5. They fully confirm but do not significantly improve the constraints obtained from the S2 orbit alone. The orbital periods in the central arcsecond range between 15 and a few hundred years. Most of the orbits appear to have a moderate to high eccentricity ('radial' orbits). Using the orbital accelerations of four of the six orbits in Figure 6, Schödel et al. find from a maximum-likelihood analysis that the center of gravitational force on all orbits is at the radio position of SgrA* to within the combined uncertainty of 20 mas (Figure 6, right inset).

The Keck group very recently have reported their first orbit analysis of the SgrA* cluster stars as well (Ghez, priv. comm., Ghez et al. 2003). Their results are in excellent agreement with ours, and in some cases improve the constraints significantly. Particularly exciting is the star S0-16, which has a very high excentricity (0.97) and approached SgrA* in late 1999 to within about 11 light hours (see also Fig.6). Ghez et al. (2003) were also able to extract a radial velocity of S2 from detection of blue-shifted Brγ absorption. This resolves the 180° inclination ambiguity of the orbit and shows that the angular momentum of S2's orbit is counter to that of Galactic rotation.

## Constraints on the Nature of SgrA*

The new Galactic Center data prove beyond any doubt that the gravitational potential resembles that of a point mass down to a scale of 10 light hours. All available data are very well fit by the superposition of a $2.6(\pm0.2)\times10^6$ $M_\odot$ point mass, plus the extended mass distribution of the visible stellar cluster. The contribution of the extended stellar cluster around SgrA* to the total mass cannot be more than a few hundred solar masses within the peri-center distance of the orbit of S2. If the point mass is replaced by a hypothetical extended (dark cluster) mass distribution, its density must exceed $10^{17}$ $M_\odot pc^{-3}$, more than ten orders of magnitude greater than the density of the visible cluster at ≥1" from the center. In addition, such a dark cluster must have a well defined surface with a very steep density drop-off outside of it, in order to fit the flat mass distribution over three orders of magnitude in radius. If such a hypothetical dark cluster consists of low mass stars, neutron stars, or stellar black holes, the inferred density implies a lifetime less than a few $10^5$ years (Maoz 1998). This is obviously a highly improbable configuration. All stars in its vicinity are much older. Theoretical simulations of very dense, core collapsed clusters predict much shallower, near isothermal density distributions, inconsistent with the flat mass distribution. Such a dark cluster model can now be safely rejected. Our new data also robustly exclude one of two remaining, 'dark particle matter' models as alternatives to a supermassive black hole, namely a ball of heavy (10-17 keV/c²) fermions (sterile neutrinos, gravitinos or axinos) held up by degeneracy pressure. Such a fermion ball could in principle account for the entire range of dark mass concentrations in galactic nuclei with a single physical model (Tsiklauri & Viollier 1998). Because of the finite size (~0.9") of a $3\times10^6$ $M_\odot$ ball of ~16 keV fermions, the maximum (escape) velocity is about 1700 km/s and the shortest possible orbital period for S2 in such a fermion ball model would be about 37 years, clearly inconsistent with the orbits of S2 and S0-16. The enclosed mass at the peri-centers of S2 and S0-16 would require neutrino masses of >58 and >67 keV. Apart from the fact that the existence a of massive neutrino is not favored by current neutrino experiments, the fermion ball model can now be safely excluded from the above constraints if the model is to explain the entire range of observed dark masses in galaxy nuclei with a single particle. The only dark particle matter explanation that cannot be ruled out by the present data is a ball of weakly interacting bosons. Such a configuration would have a radius only several times greater than the Schwarzschild radius of a black hole (Torres et al. 2000). However, it would be very hard to understand how the weakly interacting bosons first manage to reach such a high concentration (Maoz 1998), and then avoid collapsing to a black hole afterwards by baryonic accretion (Torres et al. 2000). Such accretion definitely occurs in the Galactic Center at a rate of $10^{-7}$ to $10^{-5}$ $M_\odot$ $yr^{-1}$.

Figure 7 summarizes all current constraints on the size and density of the dark mass in the Galactic Center. In the last 15 years, gas and stellar dynamics have definitely proven the existence of a dark, non-stellar mass concentration of 3 million solar masses. Over this time period the observational constraints have pushed its size downward by 2.5 orders of magnitude, and its density upward by 9 orders of magnitude. Radio VLBI (Doeleman et al. 2001) and X-ray observations (Baganoff et al. 2001) have shown the existence of hot and relativistic gas on scales of ten Schwarzschild radii at the center of this mass concentration. VLA/VLBA data constrain the proper motion of SgrA* in the Galactic Center frame to be less than 20 km/s (Reid et al. 1999). Comparison of this limit to the $>10^3$ km/s velocities of the stars orbiting SgrA* then implies a lower bound to the mass of the radio source of $10^{3...4}$ $M_\odot$, and to its density of $10^{19}$ $M_\odot$ $pc^{-3}$ (Reid et al. 1999, Chatterjee et al. 2002). In the allowed upper left corner of parameter space there remain two possible configurations, one of which (the boson star) is purely speculative and also highly unstable. It thus seems safe to conclude that SgrA* must be a 3 million solar mass black hole, beyond any reasonable doubt. The black hole appears to be fairly 'naked' (see the constraints above on any additional extended emission surrounding it). It could in principle be a tight (<10 light hour separation) binary black hole. However, an equal mass, binary hole would coalesce by gravitational radiation in a few hundred years. The observation of radial velocity anisotropy in the cluster of stars closest to SgrA* (Schödel et al. 2003) appears also to favor a single black hole: A strong tangentially anisotropic velocity dispersion would be expected near a binary black hole. The Galactic Center thus has presented us with the best astrophysical evidence we now have that the objects predicted by Einstein's and Schwarzschild's theoretical work are in fact realized in the Universe.

While the first year of NAOS/CONICA observations have led to an unexpected breakthrough in the studies of orbits, not much has been learned as yet about the emission properties of SgrA* itself. The very fact that the bright star S2 was so close in 2002 to the central source has prevented meaningful observations of its emission. The deduced upper limits to SgrA*'s flux in the H, $K_s$ and L' bands are somewhat lower than previous observations but are broadly consistent with current low accretion CDAF, ADIOS or jet models of the source (Melia & Falcke 2001).

## Outlook

The NAOS/CONICA work we have presented here is just the first step in a new phase of near-infrared observations of the immediate surroundings of the central dark mass at the nucleus of the Milky Way. The observation of Keplerian orbits offers a clean new tool of high precision gravitational studies. Stellar orbits also allow an accurate determination of the distance to the Galactic Center. With CONICA/NAOS we are presently able to distinguish more than 30 individual stellar sources within 0.5" (~23 light days) from SgrA*. For a number of them we hope to determine high precision orbits in the next few years. Some of them may approach the black hole within a few light hours, with velocities approaching 10% of the speed of light and orbital time scales of less than a few years. In addition to imaging, spectroscopy will reveal the properties of the stars bound to the hole, and also give additional information on the gravitational potential and on interactions between stars. Time resolved photometry and polarimetry will allow studies of the accretion flow and gravitational lensing effects. Another order of magnitude improvement in spatial resolution will become possible with infrared interferometry at the VLTI (once the PRIMA facility is available), the Keck interferometer and the Large Binocular Telescope. These observations will provide a few to 10 mas (a few light hours) resolution and offer exciting prospects for the exploration of relativistic motions at 10-100 Schwarzschild radii from the central black hole.

*Acknowledgments.* *We are grateful to the ESO Director General, Catherine Cesarsky, and the Director of Paranal Observatory, Roberto Gilmozzi for making these observations possible. We also thank the NAOS and CONICA team members for their hard work, as well as the staff of Paranal and the Garching Data Management Division for their support of the commissioning and science verification. We are grateful to B.Schutz for a discussion of the lifetime of a hypothetical binary black hole.*

**Figure captions**

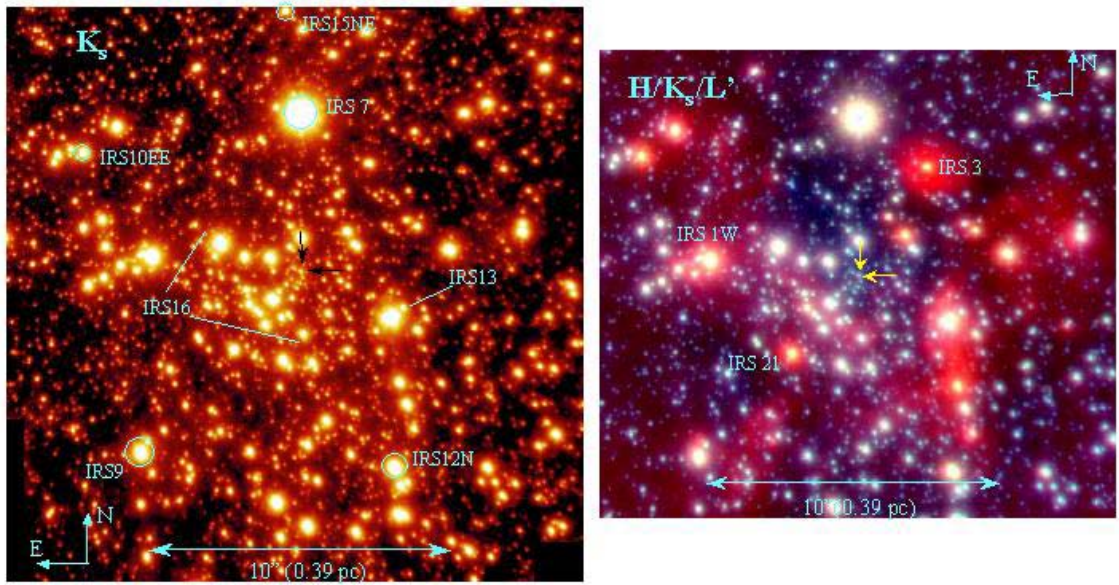

Figure 1. Left: $K_s$-band NAOS/CONICA shift-and-add image of the Galactic Center taken in August 2002 (55 mas FWHM resolution, Strehl ratio >50%: from Schödel et al. 2002, Genzel et al. 2003). East is to the left, and north is up. The brightest star, IRS7 (6.7 mag) was used as infrared AO guiding star. The five encircled stars are also radio SiO masers and have been used for establishing the astrometry (Reid et al. 2003). The IRS16 and IRS13 complexes of emission line stars are marked. The two arrows denote the position of SgrA*. Right: H/$K_s$/L' three color composite of the central ~20″. Several bright, dusty L-band excess stars are marked. The two arrows denote the position of SgrA*.

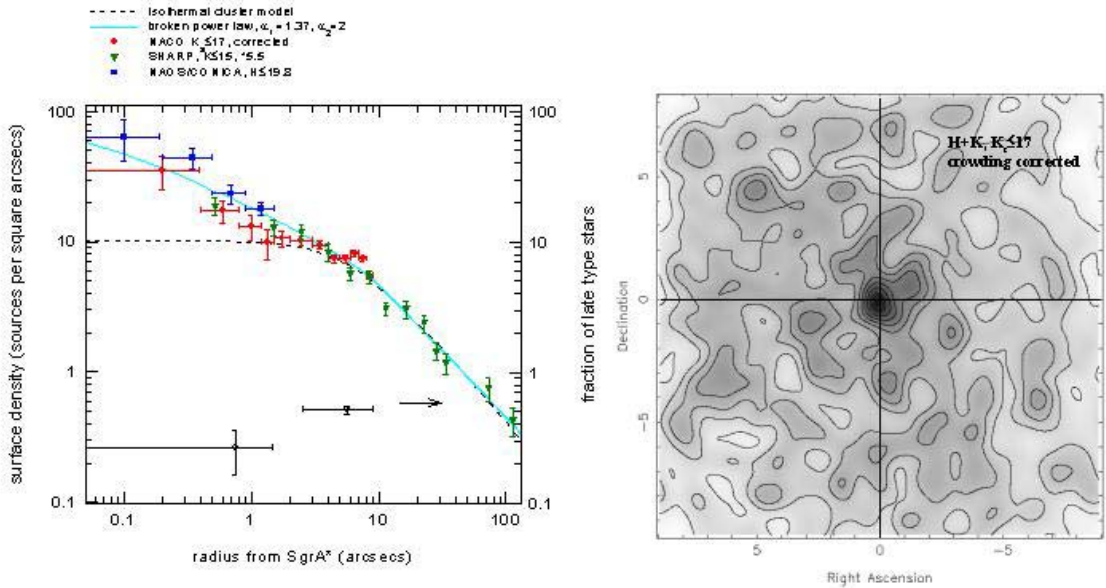

Figure 2. Surface density distribution of stars (from Genzel et al. 2003). Left: Surface density of stars as a function of projected radius from SgrA*. Red filled circles are NAOS/CONICA counts (incompleteness corrected) of sources with $K_s \leq 17$. Filled blue squares denote direct H-band NAOS/CONICA counts in the central region. Downward-pointing filled green triangles denote scaled SHARP $K \leq 15$ star counts. The short-long dash curve is the model of an isothermal sphere of core radius 0.34 pc fitting the outer SHARP counts. The blue continuous curve is the broken power-law ($\alpha=2$ beyond 10" and $\alpha=1.37$ within 10″). Open blue circles at the bottom of the figure denote the fraction of late type stars of the total $K \leq 15$ sample with proper motions and Gemini CO narrow-band indices. All vertical error bars are ±1σ, and include the statistical and incompleteness correction error. Horizontal bars denote the width of the annulus. Right: Two-dimensional map of the smoothed surface density of NAOS/CONICA incompleteness corrected source counts. Contours are 10, 20,......90, 95, 99 and 99.9% of the peak surface density. The maximum of the stellar density is at (RA,Dec)=(0.09", -0.15") relative to SgrA*, with an uncertainty of ±0.2".

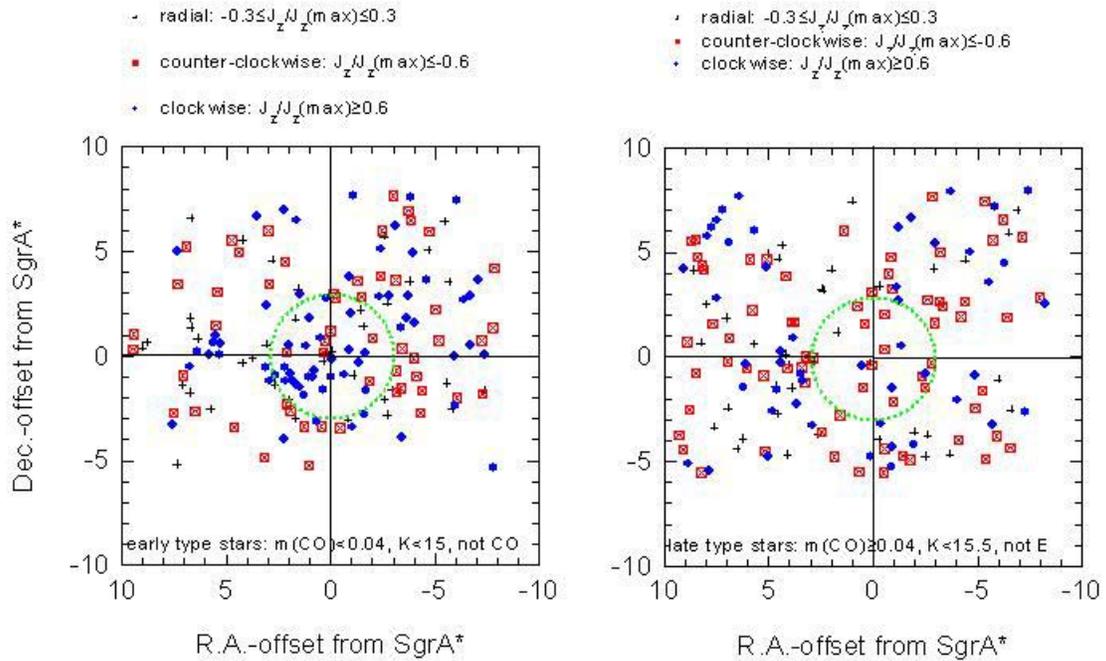

Figure 3. Spatial distribution of stars with different line-of-sight, normalized angular momenta ($J_z/J_z(max)$), obtained from the K≤15 proper motion data set of Ott et al. (2003, see Genzel et al. 2003). Stars with mainly tangential, clockwise motions on the sky are marked as filled blue circles, stars with tangential, counter-clockwise motions are marked as crossed, red open squares, while stars with motions mostly along the projected radius vector are marked as grey crosses. Keep in mind that these are sky-projected motions. The right inset shows the distribution of spectro-photometrically identified, late type stars, while the left inset shows the same for early type stars.

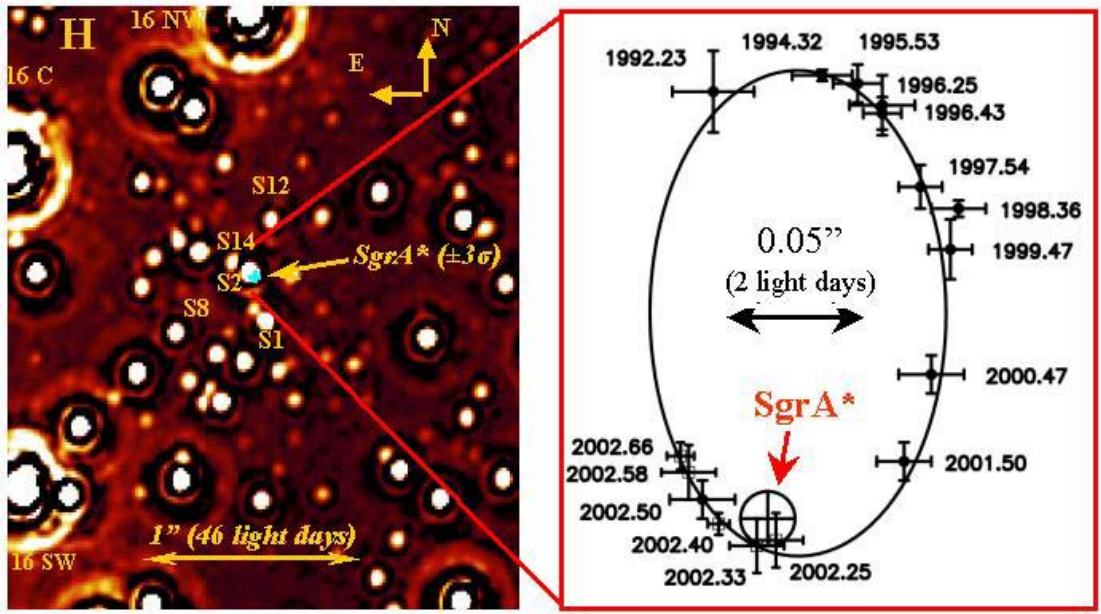

Figure 4. Orbit of the star S2 around SgrA* (from Schödel et al. 2002, 2003). Left: Wiener filtered, H-band NAOS/CONICA image (40 mas resolution) of the central 2". The light blue cross denotes the position of SgrA*, and its 3σ error bar. Several of the blue 'S'-stars in the SgrA*-cluster are marked, as well as three of the bright IRS16 stars. The right inset shows the orbital data and best keplerian fit of S2 around the position of Sgr* (circle with cross). Filled circles are measurements with the SHARP speckle camera on the 3.5m NTT, and open squares are the new NAOS/CONICA measurements. Error bars are conservative estimates, including systematic uncertainties.

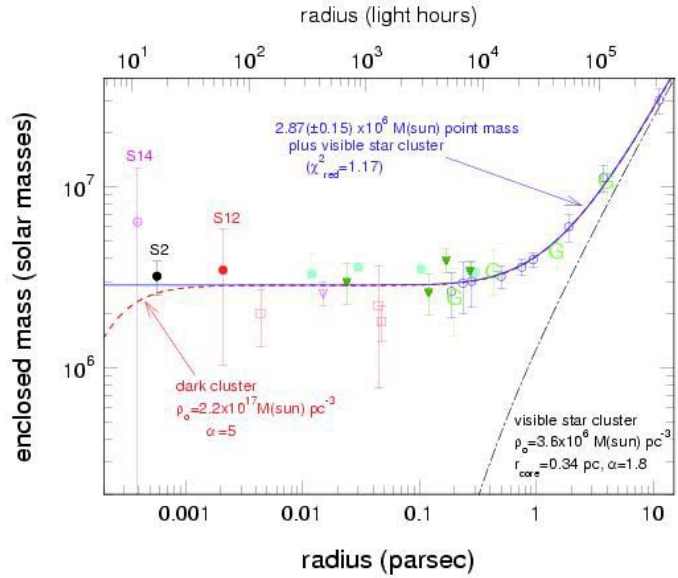

Figure 5. Derived mass distribution in the Galactic Center (for an 8 kpc distance: Schödel et al. 2002, 2003). The filled black circle denotes the mass derived from the orbit of S2, the pink filled square the mass derived from the orbit of S3, and the filled green circle the mass derived from the orbit of S0-16 (Schödel et al. 2003). Error bars combine statistical (fit) and systematic/astrometry errors. Filled dark green triangles denote Leonard-Merritt projected mass estimators from a new NTT proper motion data set by Ott et al. (2003), separating late and early type stars, and correcting for the volume bias in those mass estimators by scaling with correction factors (0.88-0.95) determined from Monte Carlo modeling of theoretical clusters. An open up-pointing triangle denotes the Bahcall-Tremaine mass estimate obtained from Keck proper motions (Ghez et al. 1998). Light-blue, filled rectangles are mass estimates from a parameterized Jeans-equation model, including anisotropy and differentiating between late and early type stars (Genzel et al. 2000). Open circles are mass estimates from a parameterized Jeans-equation model of the radial velocities of late type stars, assuming isotropy (Genzel et al. 1996). Open red rectangles denote mass estimates from a non-parametric, maximum likelihood model, assuming isotropy and combining late and early type stars (Chakrabarty & Saha 2001). Black letter 'G' points denote mass estimated obtained from Doppler motions of gas (see Genzel & Townes 1987 and references therein). The black continuous curve is the overall best fit model to all data. It is a sum of a $2.6(\pm 0.2) \times 10^6$ $M_\odot$ point mass, plus a stellar cluster of central density $3.9 \times 10^6$ $M_\odot/pc^3$, core radius 0.34 pc and power-law index $\alpha=1.8$. The grey long dash-short dash curve shows the same stellar cluster separately, but for a infinitely small core (i.e. a 'cusp'). The thick dashed curve is a sum of the visible cluster, plus a Plummer model of a hypothetical concentrated ($\alpha=5$), very compact ($R_0=0.00019$ pc) dark cluster of central density $1 \times 10^{17}$ $M_\odot pc^{-3}$.

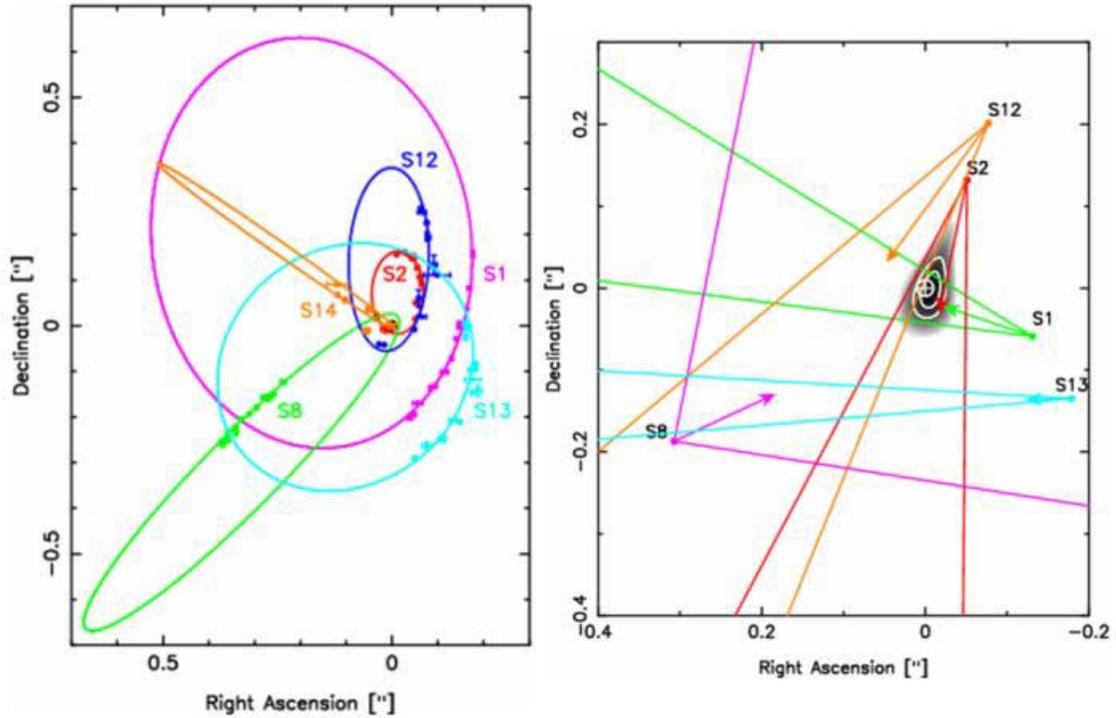

Figure 6. Stellar orbits in the SgrA*-cluster (Schödel et al. 2003). Left inset: Stellar orbits of 6 stars, obtained from an analysis of the 10 year SHARP-NAOS/CONICA data set. These innermost stars are on fairly elliptical orbits, with orbital periods between 15 and a few hundred years. The right inset shows a comparison of the best maximum-likelihood distribution of the center of gravitational force determined from the S1, S2, S3 and S8 orbits (grey scale, with 1,2,3 $\sigma$ contours), with the nominal radio position of SgrA* (white cross with 1$\sigma$ error circle). Radio position and center of force agree to within 20 mas.

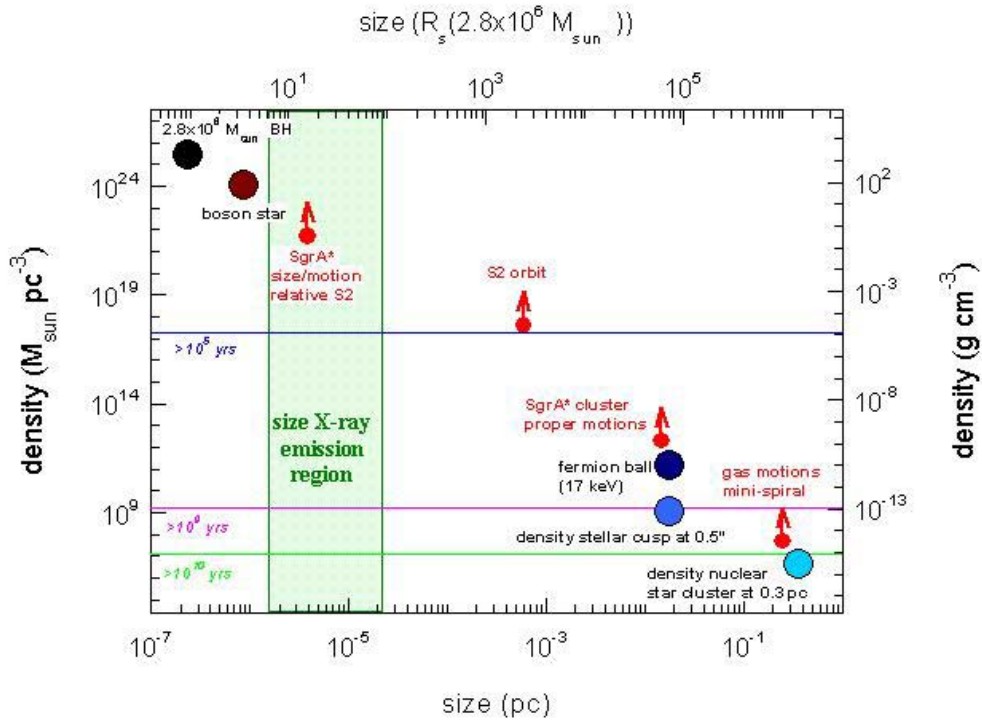

Figure 7. Constraints on the nature of the dark mass concentration in the Galactic Center (from Schödel et al. 2003). The horizontal axis is the size (bottom in parsec, top in Schwarzschild radii), and the vertical axis density (left in solar masses per pc$^3$, right in g cm$^{-3}$). Red filled circles denote the various limits on the size/density of the dark mass discussed in this paper. In addition the grey shaded area marks the constraints on the size of the variable X-ray emission from Baganoff et al. (2001). Large filled circles mark the location of different physical objects, including the visible star cluster and its central cusp, the heavy neutrino, fermion ball of Tsiklauri & Viollier (1998), the boson star of Torres et al. (2000) and, in the top left, the position of a 2.6x10$^6$ M$_\odot$ black hole. In addition three thin horizontal lines mark the lifetimes of hypothetical dark clusters of astrophysical objects (neutron stars, white dwarfs, stellar black holes, rocks etc., Maoz 1998).